\documentclass[aps,prd,floatfix,superscriptaddress,preprintnumbers,twocolumn,longbibliography]{revtex4-1}

\usepackage{amsmath}
\usepackage{amssymb}    
\usepackage{amsfonts}
\usepackage{graphicx}   
\usepackage{verbatim}   
\usepackage{color}      
\usepackage{subfigure}  
\usepackage{hyperref}   
\begin{document}

\title{Gravitational Fermion Creation During An Anisotropic Phase Of Cosmological Expansion}
\author{Amit Bhoonah} 
\email{amit.bhoonah@queensu.ca}\begin{flushleft}
•
\end{flushleft}
\affiliation{The McDonald Institute and Department of Physics, Engineering Physics,
and Astronomy, Queen’s University, Kingston, Ontario, K7L 2S8, Canada} 
\affiliation{Eidgen{\"o}ssische Technische Hochschule Z{\"u}rich, R{\"a}mistrasse 101, 8092 Z{\"u}rich, Switzerland}
\date{\today}

\begin{abstract}
The free Dirac equation is solved in a Bianchi Type I space-time, which represents a homogeneous but anisotropic universe, to show the creation of fermionic particles. It is found that unlike in the isotropic case, massless fermion production is possible. An estimate of the energy density of massless particles created during an early anisotropic phase of cosmological expansion is shown to cause substantial back-reaction on the gravitational field. The potential relevance to dark matter particle production, primordial magnetogenesis, and early universe cosmology is discussed briefly. 
\end{abstract}

\maketitle

\section{Introduction}\label{sec:intro}
The quantisation of matter fields in curved space-time is a rich subject that has been treated extensively since the late 1960s. In \textit{semi-classical} gravity, or Quantum Field Theory (QFT) in curved space-time as it is commonly called, the gravitational field is classical and treated using General Relativity while fermionic, gauge, and scalar fields (collectively called matter fields in this work) are quantized and treated as a QFT. The curved background gives rise to a rich variety of phenomenology not present in flat space-time, such as the creation of particles by the gravitational field. This is a consequence of energy not being conserved in the most general gravitational field. There is no globally defined vacuum for the entire space-time in the manner one is familiar with from the treatment of free fields in QFT: a state of lowest energy containing no particles. Two different vacuua at two different space-time points will only be empty with respect to each other in a limited number of scenarios, representing a major difficulty in the mathematical setup of the problem. A typical approach in the literature has been to assume that the space-time under consideration is asymptotically flat and define vacuum states in those regions. These $in$ and $out$ vacuua are \textit{inequivalent} and each one contains particles relative to the other - the gravitational field has created particles. \\

This phenomenon of particle creation by gravitational fields was first shown by Parker in \cite{parkerparticlecreation}, \cite{parkerspin0}, and \cite{parkerspin12}. He and several others afterwards (see for example \cite{tagirov}) studied the quantisation of matter fields in isotropic and homogeneously expanding universes described by the Friedmann-Le Ma\^itre-Robertson-Walker (FLRW) metric. It was found, as might seem intuitive, that the number density of created particles decreases at higher masses. One therefore expects that low or zero mass particle production occurs at a rate quite high compared to the more massive case, motivating the study of gravitational particle creation in the massless limit. \\

However, a major result in FLRW  universes is that massless fermions or vector bosons - such as photons - are \textit{not} created gravitationally. The reason has to do with conformal invariance. For particles of non-zero spin the mass term is the only one that breaks conformal symmetry. Therefore, in the massless fermion or vector boson limit a conformal transformation can be used to recast the equations of motion in an FLRW space-time into their Minkowskian forms, in which there is no gravitational particle creation. \\

The situation is slightly more complicated for spin-0 particles. The most general Lagrangian that can be written on account of renormalizability includes a coupling between the Ricci scalar $\mathcal{R}\left(x\right)$ and the scalar field $\phi$: 
\begin{equation}\label{eq:scalarlagrangian}
\mathcal{L} = \frac{1}{2}\sqrt{-g}\left[g^{\mu\nu}\partial_{\mu}\phi\partial_{\nu}\phi - \frac{m^2}{2}\phi^{2} - \xi\mathcal{R}\phi^{2}\right],
\end{equation}
where $g^{\mu\nu}$ is the metric with determinant $g$, $\xi$ is a dimensionless parameter, and a quartic interaction has been omitted since only terms quadratic in the scalar field are being considered. Two values of $\xi$ are of particular interest: the minimally coupled case, $\xi = 0$, and the conformally coupled case, $\xi = \frac{1}{6}$, for which~\eqref{eq:scalarlagrangian} is conformally invariant when the scalar is massless \cite{parkerqftcurvedspacetime}. Therefore, if $\xi = \frac{1}{6}$ massless scalars are not gravitationally created in an FLRW universe, as was shown explicitly in \cite{tagirov}. However, there is no \textit{a priori} reason for setting $\xi = \frac{1}{6}$, and for complete generality one should take $\xi$ to be an arbitrary dimensionless parameter. Of course this comes at the expense of conformal invariance for the Lagrangian, meaning that massless scalars will be gravitationally produced.  \\    
 
There is a subtlety in the above discussion of conformal invariance that deserves further mention. That one can perform a conformal transformation which reduces the equations of motion for massless fermions, vector bosons, or conformally coupled scalars into their flat space-time forms is a property particular to the FLRW metric. The same will not be true of other metrics which are not conformally flat; even if the equations of motion are conformally invariant, it will not be possible to reduce them to their Minkowskian forms with a conformal transformation. This was illustrated by Zeldovich and Starobinsky for conformally coupled scalar fields in a Bianchi Type I universe, which is spatially homogeneous but not isotropic \cite{bianchi1spin0}.  Anisotropy is achieved by allowing the scale factors of the three spatial directions to differ and homogeneity is preserved by restricting the metric elements to functions of time but not space. \\

The central result of \cite{bianchi1spin0} is that conformally coupled massless scalars are created in Bianchi Type I universe but that this stops in the isotropic limit. In this current work I extend this result to spin-$\frac{1}{2}$ fields. It is shown that massless fermions are gravitationally created in a Bianchi Type I space-time but that this stops in the isotropic limit, as one might anticipate from the preceeding discussion of conformal invariance.  \\

Before discussing the motivations for studying the creation of massless fermions in an anisotropically expanding space-time, a brief cosmological detour is apropos. Particle production in an FLRW space-time is particularly relevant when the scale factor is increasing exponentially, for example in an inflationary scenario. Massless minimally coupled scalars are abundantly produced during inflation and considered to be the main source of scalar perturbation while gravitons, produced because the Einstein-Hilbert action is not conformally invariant, are the source of tensor perturbations (the interested reader is invited to consult R. P. Woodard's extensive review for a deeper discussion of the subject \cite{woodard}). Clearly, standard cosmology, which is well established experimentally, offers useful mechanisms for the gravitational creation of massless particles and one does not have to go to non-standard scenarios like an anisotropic phase of expansion to obtain such results. Any massless fermions or vector bosons could be produced by the subsequent decay of the gravitationally produced minimally coupled scalars. However, this would only apply to particles that couple to the scalar and even then such a process might be highly suppressed. \\

The main motivation behind this work is that the gravitational creation of massless fermions or vector bosons during an anisotropic phase of cosmological expansion offers an additional way of producing these particles. This paper starts with a brief review of the mathematical setup of semi-classical gravity in Section~\eqref{sec:intro2}, describing in particular the conditions that determine the restricted class of space-times for which it is possible to define a vacuum state globally. Particular attention is paid to the role of conformal invariance. In Section~\eqref{sec:intro3}, it is argued that the equivalence principle allows the definition of local vacuua around any arbitrary point in the gravitational field provided some reasonable requirements are met. This has the benefit of doing away with the asymptotic flatness assumption and being applicable to more general situations of physical interest. The main result of the paper - the creation of massless fermions in a Bianchi Type I universe - is shown in Section~\eqref{sec:canonquant}. Since any cosmological model must explain the observed homogeneity and isotropy of the universe, it is tempting to believe that the effects of particle production during a very early anisotropic phase of the expansion will have negligible influence on the evolution of the universe. However, in Section~\eqref{sec:conc} a coarse estimate of the energy density of created massless fermions is found to be substantial for times close to but not at or before the Planck time. This is the same conclusion reached for scalar fields in \cite{bianchi1spin0}, although I do not attempt a full computation of the energy momentum tensor. Three examples where massless particle creation during an early anisotropic phase might be relevant are also mentioned as concluding remarks. The first two are that such a phase provides a mechanism for dark matter production and primordial magnetic field generation. For the third example, it is argued that the substantial amount of created massless fermions around the Planck time deserves further investigation as it \textit{may} alter the form of collapse to a singularity of a hypothetical cosmological contraction phase without stopping it. This is similar to what was found in \cite{bianchi1spin0} for scalar fields.  \\
  
\section{Quantisation In Curved Space-Time}\label{sec:intro2}
\subsection{Killing Vector Fields and Energy}
Quantisation in the presence of a classical gravitational field is severely plagued in one major way: in curved space-time Poincar\'e  symmetry is broken. This is problematic because when constructing QFTs, mostly gauge theories motivated by dark matter and other phenomena that the Standard Model of Particle Physics cannot explain, Poincar\'e invariance - a term often used interchangeably with Lorentz invariance in the literature - is assumed \textit{de facto}. This allows for particles and anti-particles to be fundamental concepts in QFT. A particle or an anti-particle is an excitation of the quantised matter field that produces a localised quanta of energy, particles having positive energy and anti-particles having negative energy. What enables this unambiguous definition of particles and anti-particles is the fact that in flat space-time energy is a well defined and conserved quantity. The quantised fields can be written as plane waves of negative (particles) and positive (anti-particles) frequencies representing positive and negative energies respectively. This is in turn related to the fact that the Poincar\'e group admits time translation as a symmetry - energy being the associated conserved quantity. Intuitively, one may infer that a space-time - curved or flat - that is invariant under time translations (meaning energy is conserved) will have a fundamental notion of particles and anti-particles, even if the isometries of the space-time in question do not correspond to the full Poincar\'e group. The problem at hand can hence be formulated as a two part question:  \textit{under what conditions is a space-time invariant under time translations and how does one construct a QFT if such conditions are not met?} \\

The answer to the first part of this question is of course well known and the material presented in this section is intended as a brief summary to motivate the approach taken in this work to address the second part. The material presented here draws heavily from the classic texts \cite{dewitt}, \cite{birrell}, and \cite{parkerqftcurvedspacetime} and the interested reader is invited to consult these references for a more detailed treatment of the subject, or \cite{blureview} for a more recent review. \\

A space-time with a metric $g_{\mu\nu}$ and covariant derivative \small $\nabla_{\mu} = \partial_{\mu} + \Gamma_{\mu\nu}^{\rho}$ ($\Gamma_{\mu\nu}^{\rho}$ \normalsize are the Christoffel symbols of the metric, with space-time labels suppressed for clarity - this will be done throughout this text) is invariant with respect to time translations if it possesses a time-like Killing vector, i.e., a vector $K$ satisfying
\begin{align}
\nabla_{\mu}K_{\nu}  + \nabla_{\nu}K_{\mu} = 0 \label{eq:killingvec} \\
g^{\mu\nu}K_{\mu}K_{\nu}  > 0 \label{eq:timelikevec},
\end{align} 
where~\eqref{eq:killingvec} is the condition that the Lie derivative of the metric with respect to the vector $K$ vanishes. When such a time-like Killing vector exists, there will also be a time-like coordinate, denoted by $t$ here, and it will be possible to choose a coordinate system in which the metric is independent of that coordinate. When setting up a coordinate system for the corresponding gravitational field, $t$ is therefore a \textit{natural} choice to represent time. It is important to highlight here that conditions~\eqref{eq:killingvec} and~\eqref{eq:timelikevec} are completely general in the sense that no assumption has been made about whether the space-time is flat or curved. In the case of flat space-time, the covariant derivatives simply reduce to partial derivatives as the connections vanish. \\
\\
If the theory is conformally invariant, condition~\eqref{eq:killingvec} can be relaxed to that of $K$ being a conformal killing vector, 
\begin{equation}
\nabla_{\mu}K_{\nu}  + \nabla_{\nu}K_{\mu} = \lambda\left(x\right)g_{\mu\nu}, \label{eq:conformalkillingvec}
\end{equation}
with $\lambda\left(x\right)$ a scalar function of space-time. For non-interacting fields the mass term is the only source of conformal symmetry breaking, with the exception of scalar fields with a non-conformal coupling to the Ricci scalar \cite{parkerspin0}. A space-time containaing only massless non-interacting fields will therefore be invariant under time-translations if it possesses a time-like conformal Killing vector. \\

The Killing vector $K$ defines energy as the conserved quantity $E$,
\begin{equation}\label{eq:energy}
E = \int_{\Sigma} K_{\mu}T^{\mu\nu} d\Sigma_{\nu}
\end{equation}
where repeated indicies are summed over, \small $T^{\mu\nu}$ \normalsize is the Energy Momentum Tensor, \small $\Sigma$ \normalsize is a Cauchy hypersurface, and $E$ is independent of the choice of \small $\Sigma$. \normalsize
\subsection{Field Decomposition, Natural Bases, and the G Vacuum}
Having adopted a coordinate system with the \textit{natural} time and found a globally acceptable definition of energy, one now proceeds with the quantisation of matter fields existing on that background space-time. The matter fields, denoted generically by $\varphi\left(x\right)$ here, obey the Euler-Lagrange equations of motion obtained from the Lagrangian, and, in seeking solutions to these equations, one considers a basis for the decomposition of the solutions into negative and positive frequencies, corresponding to particles with positive energies and and anti-particles with negative energies respectively. There is no unique way of doing this. However, with the time-like Killing vector $K$ and coordinate system with the metric independent of $t$, there is \textit{natural} eigenfunction basis $\mathit{e}_j$,
\begin{equation*}
\begin{split}
\varphi_{j}\left(x\right) & \propto a_{\mathbf{\omega_{j}}}\mathit{e}_{j} + a_{\mathbf{\omega_{j}}}^{\star}\mathit{e}^{\star}_{j} \\
\mathcal{L}_{K}e_{j} & = - i \omega_{j}e_{j} \\
\mathcal{L}_{K}e^{\star}_{j} & = + i \omega_{j}e^{\star}_{j}
\end{split}
\end{equation*}
where $j$ is a generic label for the quantum numbers used to describe the field and the full solution is a super-position of eigenfunctions \small $\varphi_{j}$. $\mathcal{L}_{K}$ \normalsize is the Lie derivative of the field with respect to the vector $K$ and $\omega_{j}$ is the energy of the state $e_{j}$. In the \textit{natural} basis for the fields and with \small $K = \left(1,0,0,0\right)$ \normalsize - $K$ is written using the basis \small $V_{\mu} = V^{\mu\nu}\partial_{\nu}$ \normalsize for vectors - this reduces to  
\begin{equation*}
\frac{\partial e_{j}}{\partial t} = - i \omega_{j}e_{j},
\end{equation*}
with  $\mathit{e}^{\star}_{j}$ satisfying the same equation but with a $+ i$ sign. It should again be emphasized that this discussion is completely general and includes the flat space-time case. The solutions to the above equation are the exponentials, $\varphi_{j}\left(x\right) \propto \exp\left(\mp \ i\omega_{j}t\right)$, one is familiar with from QFT. \\

After imposing canonical quantisation on the fields, the coefficients $a_{\mathbf{\omega_{j}}}$ and $a_{\mathbf{\omega_{j}}}^{\star}$ of the basis functions $e_{j}$ and their conjugates $e_{j}^{\star}$ respectively will be interpreted as the creation and annihilation operators of quantas of energy $\pm \ \omega_{j}$. The energy $E$ being a conserved quantity, a vacuum state can now be defined
\begin{equation}
E \ \vert \ 0 \ > \ = \ 0
\end{equation}
In anticipation of the next section, I will call this vacuum a global, or $G$, vacuum as it is unique on the entire manifold up to isometries of the space-time. It is a \textit{true} vacuum in the sense that it can be unambiguously stated that it contains no particles. The Fock space is constructed by repeated application of creation operators on the vacuum just like in QFT. 

\section{The General Case}\label{sec:intro3}
The discussion until now assumed that the space-time under consideration possesses a global time-like Killing vector. This is a rather restrictive. A general space-time need not possess any Killing vectors \textit{at all}, let alone a time-like one. This is a serious challenge to the foundation upon which QFTs are built and, before proceeding further, it is helpful to understand the full range of scenarios that one may encounter. DeWitt gave a comprehensive account of these in \cite{dewitt}:
\begin{enumerate}
\item \label{one}There may be no global Killing vector at all, time-like or otherwise. In the literature, this has been typically dealt with by assuming that the space-time is asymptotically flat. Only in those limits can a meaningful definition be assigned to particles and anti-particles. The set up includes two asymptotic and inequivalent \textit{IN} and \text{OUT} vacuua with particle creation between the two. This may not be very applicable to most physical situations of interest - especially in the cosmological context - and one of the chief aims of this paper is to show that more general assumptions are possible.
\item \label{two}Even if there exists a time-like Killing vector, it may not be time-like everywhere on the manifold. A typical example is the Schwarzchild metric, for which $K = \left(1,0,0,0\right)$ is time-like outside the event horizon but space-like inside it. 
\item \label{three} The space-time may admit a time-like Killing vector only within a limited region. In this case, there are several vacua - one for every region in which there exists a time-like Killing vector. DeWitt argued in favour of a principle of relativity by which all of these vacua must be equivalent \cite{dewitt}. This is the principle that will be used in this work for the quantisation of matter fields in curved space-time. Using the nomenclature of Chernikov and Tagirov \cite{tagirov}, these will be called quasi, or $Q_{X}$, vacua, with $X$ the region of space-time in which the vacua are defined.
\end{enumerate}

The quasi vacuum in is in fact \textit{guaranteed} for any arbitrary metric by virtue of the equivalence principle. At an arbitrary point in a gravitational field, a locally inertial coordinate system may be constructed for a small neighbourhood around that point. One must however be careful to avoid situations where the small neighbourhood includes a singularity or a horizon. To steer clear of such situations, I will consider only \textit{ordinary} points, by which I mean that the small neighbourhood around them contains neither a singularity nor a horizon. Within that small neighbourhood, a QFT may be set up. \\

Naturally, one now needs to precisely define the small neighbourhood within which the QFT is set up. To that end, it is helpful to think of the Christoffel symbols as spurions that break Poincar\'e symmetry. Returning to the Killing equation and writing it out explicitly, 
\begin{equation*}\label{eq:killingeqexplicit}
\partial_{\mu}K_{\nu} + \partial_{\nu}K_{\mu} + 2\Gamma_{\mu\nu}^{\rho}K_{\rho} = 0.
\end{equation*}
In the limit of vanishing Christoffel symbols, full Poincar\'e symmetry is recovered. The magnitude of \small $\Gamma_{\mu\nu}^{\rho}\left(x\right)$ \normalsize therefore indicates the degree to which Poincar\'e symmetry is broken. On the scale of quantum processes in weak gravitational fields, this can be safely ignored. \\ 

The size of the time translation spurion is of particular interest. To obtain this, one first fixes an \textit{ordinary} point $X$ as the reference point and considers a nearby point $x$. Using the notation of Appendix~\eqref{sec:Tetradform}, the tetrads \small $e^{\bar{\mu}}_{\nu}\left(X\right)$ \normalsize can be used to define a local orthonormal basis \small $\tilde{\partial}_{\bar{\mu}}=e^{\nu}_{\bar{\mu}}\left(X\right)\partial_{\nu}$ \normalsize for vectors. Applying the Killing condition to the vector \small $K_{X} = \left(e^{\nu}_{\bar{0}}\left(X\right)\partial_{\nu}, 0, 0, 0\right)$ \normalsize and using the expansion in terms of local inertial coordinates described in Appendix~\eqref{sec:Tetradform},
\begin{equation*}
\nabla_{\mu}K_{\nu} + \nabla_{\nu}K_{\mu} = -\frac{2}{3}\left(\tilde{R}^{0}_{\mu\nu\rho} + \tilde{R}^{0}_{\nu\mu\rho}\right)\Delta X^{\rho} + \mathcal{O}\left(\Delta X^{3}\right)
\end{equation*}
where $\Delta X^{\rho} =\left(x^{\rho} - X^{\rho}\right)$. The condition for $K_{X}$ to be a local time-like Killing vector is
\begin{equation}\label{eq:killingcondition}
\left\vert \left(\tilde{R}^{0}_{\mu\nu\rho} + \tilde{R}^{0}_{\nu\mu\rho}\right)\Delta X^{\rho} \right\vert \ll 1 
\end{equation}
In a region where~\eqref{eq:killingcondition} is satisfied, a local energy, $E_{X}$, and a $Q_{X}$ vacuum can be defined as well as annililation and creation operators, $a_{\mathbf{\omega_j}}\left(X\right)$ and $a_{\mathbf{\omega_j}}^{\star}\left(X\right)$, respectively
\begin{equation}
\begin{split}
E_{X} \vert \ 0_{X} \ > \ & = \ 0 \\
a_{\mathbf{\omega_j}}\left(X\right)\vert \ 0_{X} \ > & = 0.
\end{split}
\end{equation}
The corresponding Fock space, $\mathcal{F}_{X}$, is constructed by repeated application of $a_{\mathbf{p}}^{\star}\left(X\right)$. \\

The same procedure can of course be applied at another \textit{ordinary} point $X^{\prime}$. In this case, energy eigenfunctions and creation and annihilation operators will be related by Bogoliubov transformations 
\begin{equation}
a_{\mathbf{\omega_j}}\left(X^{\prime}\right) = D_{1}a_{\mathbf{\omega_j}}\left(X\right) + D_{-1}a_{\mathbf{\omega_j}}^{\star}\left(X\right)
\end{equation}
where $D_{\pm 1}$ are the Bogoliubov matrix elements with their space-time dependence suppressed and $a_{\mathbf{\omega_j}}\left(X^{\prime}\right)$ satisfies the conjugate equation. The $Q_{X^{\prime}}$ vacuum is the state of lowest energy with respect to $E_{X^{\prime}}$ and is annihilated by $a_{\mathbf{\omega_j}}\left(X^{\prime}\right)$ 
\begin{equation*}
\begin{split}
E_{X^{\prime}} \vert \ 0_{X^{\prime}} \ > & = \ 0 \\
a_{\mathbf{\omega_j}}\left(X^{\prime}\right)\vert \ 0_{X^{\prime}} \ > &= 0
\end{split}
\end{equation*}
However, the $Q_{X}$ vacuum is not the state of lowest energy with respect to $E_{X}$,
\begin{equation*}
E_{X} \vert \ 0_{X^{\prime}} \ >  \ \neq \ 0
\end{equation*}
This can be seen by calculating the number density of $Q_{X^{\prime}}$ with respect to $X$
\begin{equation}
N_{X^{\prime}} = \ < 0_{X} \left\vert a^{\star}_{\mathbf{\omega_j}}\left(X^{\prime}\right)a_{\mathbf{\omega_j}}\left(X^{\prime}\right) \right\vert 0_{X} > = \left\vert D_{-1} \right\vert^2
\end{equation}
$N_{X^{\prime}}$ will only be zero if the space-time has a $G$ vacuum. If not, the $Q_{X^{\prime}}$ vacuum is not empty with respect to the $Q_{X}$ vacuum - the gravitational field has created particles. 

\section{A Cosmological Example}\label{sec:canonquant}
\subsection{Bianchi I Universes}\label{ssec:b1}
I now apply the result presented in the previous section to a Bianchi Type I metric of the form 
\begin{equation}
\label{eq:bianchimetric}
ds^{2} = dt^{2} - \sum_{j=1}^{3} r^{2}_{j}\left(t\right)dx_{j}^{2},
\end{equation}
where $r_{j}\left(t\right)$ are arbitrary functions of time. This describes a spatially homogeneous but non-isotropic space-time, with isotropy recovered in the limit $r_{j}\left(t\right) \rightarrow r\left(t\right) \ \forall j$ for an arbitrary function $r\left(t\right)$. The isotropic limit is of course the FLRW metric used in cosmology. \\

Time translation is obviously not a symmetry in this case so gravitational particle creation is expected. It is however useful to start by considering the time-like vector $K = \left(1,0,0,0\right)$ and seeing if there are any non-trivial conditions under which it is a conformal Killing vector. For illustrative purposes, it is useful to fix a $k$ and perform the coordinate transformation $dt^2 = r_{k}^{2}\left(\eta\right)d\eta$ so that the line element becomes\begin{equation*}
ds^{2} = r_{k}\left(\eta\right)^{2}d\eta^{2} - \sum_{j=1}^{3} r^{2}_{j}\left(\eta\right)dx_{j}^{2}.
\end{equation*}
The Lie derivative of the new metric with respect to $K$ is non-zero and does not satisfy the condition for being a conformal Killing vector. However, in the isotropic - FLRW - limit one finds 
\begin{equation*}
\begin{split}
\nabla_{\mu}K_{\nu} + \nabla_{\nu}K_{\mu} = - \frac{\dot{r}}{r} r^{2} = \frac{\dot{r}}{r}g_{\mu\nu}
\end{split}
\end{equation*}
where a dot denotes derivative with respect to $\eta$. A conformal Killing vector therefore exists in the FLRW limit, meaning that a $G$ vacuum can be defined if the QFT is conformally invariant. In other words, massless particles cannot be created in an isotropic and homogeneously expanding universe. \\
 
After using the coordinate transformation above to show the existence of a conformal Killing vector in the FLRW limit, I now revert to the original metric~\eqref{eq:bianchimetric} for the remainder of this work. The general case has no time-like Killing vector, so a $Q$ vacuum must be defined. Fixing an arbitrary \textit{ordinary} point $X=\left(T,0,0,0\right)$ on the manifold, one constructs a locally inertial frame around it. The spatial coordinates have been set to zero since the manifold admits spatial translational invariance as a symmetry. To linear order in $\vert X\vert$,~\eqref{eq:inertcoord} gives the Riemann normal coordinates  
\begin{equation}
\begin{split}
\xi_{X}\left(x\right) & = \left(t, r^{-1}_{j}\left(T\right)x^{j}\right) 
\end{split} 
\end{equation}
for a neighbouring point $x = \left(t + T, x, y, z\right)$. Here, local translational invariance was used to set $\xi^{\mu}_{X}\left(X\right)$ as the origin. Condition~\eqref{eq:killingcondition} for metric~\eqref{eq:bianchimetric} is     
\begin{equation}\label{eq:killingcondbianchi1}
t^{2}\left\vert \left(r^{-1}_{j}\left(T\right)\right)^{2} + r_{j}\left(T\right)r_{j}^{\prime\prime}\left(T\right)\right\vert^{2} \ll 1,
\end{equation} 

The quantum scale associated to a particle of mass $m$ is its Compton wavelength, below which quantum effects can no longer be ignored. One can therefore define the reduced Compton time $t_C$ as the typical time scale associated with quantum processes of that particle,   
\begin{equation}\label{eq:comptontime}
t_{C} \equiv \frac{\hbar}{mc^2},
\end{equation}
where the reduced Planck's constant, $\hbar$, and the speed of light $c$ have been made explicit. Substituting $t_C$ into~\eqref{eq:killingcondbianchi1} and calling $\mathcal{R\left(T\right)} = \left\vert \left(r^{-1}_{j}\left(T\right)\right)^{2} + r_{j}\left(T\right)r_{j}^{\prime\prime}\left(T\right) \right\vert$ for clarity (a scalar product over $j$ is implied in the absolute value), 
\begin{equation}\label{eq:comptonkilling}
\frac{\hbar^2}{m^2c^4}\left\vert \mathcal{R\left(T\right)} \right\vert^{2} \ll 1
\end{equation}
From this, two things are immediately apparent:
\begin{enumerate}
\item For massive particles, condition~\eqref{eq:killingcondbianchi1} is violated when $\mathcal{R} \gg m$, or the radius of curvature of the space-time, $\mathcal{R}^{-1}$, is comparable to or much smaller than the Compton wavelength of the particle. In the cosmological context, this implies that quantum effects due to semi-classical gravity are relevant in the very early universe near the Planck time. 
\item Gravitational particle creation is more relevant for lighter particles. In the massless limit, quantum effects cannot be ignored as~\eqref{eq:comptonkilling} is badly violated. Massless particles are therefore most relevant for gravitational particle creation.  
\end{enumerate} 

Finally, it should be mentioned that this semi-classical treatment of gravity is at best valid up until the Planck scale. At those energies, quantum gravitational effects can no longer be ignored. In the cosmological context studied here, this means that the theory is only valid for times $t$ larger than the Planck time, $t_{P}$, which can be thought of as the Compton time of the Planck mass. The validity of the semi-classical treatment can be parameterised by a dimensionless factor $\gamma \geq 1$,
\begin{equation}
t \ \geq \ \gamma \ t_{P} = \gamma \ \sqrt{\frac{\hbar G}{c^5}} \approx \gamma \times 10^{-44} \ sec
\end{equation} 
where $\gamma = 1$ is reserved for the most optimistic of theorists. 

\subsection{Fermion Creation}
I now proceed to explicitly show an important property mentioned above using spin-$\frac{1}{2}$ particles as an example: massless particle creation in the anisotropic case but not in the isotropic case. From Appendix~\eqref{sec:DiracCrst}, the Dirac equation for fermionic fields in space-time~\eqref{eq:bianchimetric} is 
\small
\begin{equation}\label{eq:diracbianchi}
\begin{split}
& \left[i\gamma_{0}\frac{\partial}{\partial t} +
i\sum_{j=1}^{3}r^{-1}_{j}\gamma_{j}\frac{\partial}{\partial x_{j}} + \frac{i}{2}\sum_{j=1}^{3}\frac{r^{\prime}_{j}}{r_{j}}\gamma_{0} - m\right]\psi
= 0.
\end{split}
\end{equation} 
\normalsize
where $^{\prime}$ denotes derivative with respect to time. I use the Dirac representation in which the $\gamma$ matrices are given by 
\begin{equation}
\gamma_{0} = \begin{pmatrix}
I_2 & 0 \\ 0 & -I_2
\end{pmatrix} \ \ \ \gamma_{j} = \begin{pmatrix}
0 & \sigma_{j} \\ -\sigma_{j} & 0
\end{pmatrix} \ \ 
\end{equation}\\
with j = 1,2,3. $I_2$ is the 2$\times$2 identity matrix and the $\sigma_{j}$s are the Pauli matrices. \\

Since the metric is diagonal,~\eqref{eq:diracbianchi} contains no mixed derivative terms and separable solutions are appropriate. The space-time is also spatially homogeneous, which means that momentum is a conserved quantity. For a fixed momentum $\mathbf{p}$, one therefore considers solutions of the form 
\begin{equation}
\label{eq:planewavesolnp}
\psi^{\alpha}_{\mathbf{p}} = \chi^{\alpha}_{b}\left(\mathbf{p},t\right)\exp\left(i\alpha \mathbf{p}\cdot\mathbf{x}\right)
\end{equation}
where $b$ labels a spinor index and $\alpha = \pm 1$ represent negative and positive frequency modes respectively.  Substituting~\eqref{eq:planewavesolnp} into~\eqref{eq:diracbianchi},
\begin{equation}\label{eq:posfreq}
\begin{split}
& i\gamma_{0}\frac{\partial\chi^{\alpha}_{b}}{\partial t} =
\left[\sum_{j=1}^{3} \alpha p_{j}r^{-1}_{j}\gamma_{j} - \frac{i}{2}V^{-1}V^{\prime}\gamma_{0} + m\right]\chi^{\alpha}_{b}.
\end{split}
\end{equation} 
The spin connection term can be absorbed into a redefinition of $\chi\left(t\right)$ by substituting solutions of the form
\begin{equation*}
\chi\left(t\right) =  \frac{1}{\sqrt{r1\left(t\right)r2\left(t\right)r3\left(t\right)}}\tilde{\chi}\left(t\right)
\end{equation*}
into~\eqref{eq:posfreq}. As this results in an overall factor that drops out of the equation, the tilde will be suppressed for convenience. From this substitution one obtains 
\begin{equation}\label{eq:posfreq2}
\begin{split}
& i\gamma_{0}\frac{\partial\chi^{\alpha}_{b}}{\partial t} =
\left[\sum_{j=1}^{3} \alpha p_{j}r^{-1}_{j}\gamma_{j} + m\right]\chi^{\alpha}_{b}.
\end{split}
\end{equation} 
In order to solve~\eqref{eq:posfreq2}, initial conditions are required. Taking these to be at the time $T = \gamma t_{p}$ at which a local inertial frame was set up in the previous section, one considers a linear superposition
\begin{equation*}
\begin{split}
& \sum_{\alpha=-1}^{1}a^{\mathbf{p}}_{\alpha s}\left(T\right) \chi^{\alpha}_{b}\left(T\right) 
\end{split}
\end{equation*}
where 
\begin{align}
\chi^{\alpha}_{b}\left(t\right) & \equiv \exp\left[-i\alpha\int_{T}^{t} dt^{\prime} \  \omega_{\mathbf{p}}\left(t^{\prime}\right)\right]u_{\alpha s}\left(\mathbf{p},t\right) \label{eq:soln} \\ 
\omega^{2}_{\mathbf{p}}\left(t\right) & = \sum_{j=1}^{3} \left(\frac{p_{j}}{r_{j}\left(t\right)}\right)^2 + m^2 \label{eq:energy}
\end{align} 
Some comments about the form of the expression in~\eqref{eq:soln} are in order. An exact analytic solution is available only for a limited number of functional forms of the scale factors $r_{j}\left(t\right)$. The exponential phase with an integral is obtained by ``squaring" the Dirac equation~\eqref{eq:posfreq2}, which yields the corresponding second order Klein-Gordon equation (hereafter referred to as KG) satisfied by each component of the spinor and writing a solution of the latter in terms of adiabatic mode functions, a WKB-like solution first described in \cite{parkerspin0} and also used in \cite{bianchi1spin0}, \cite{parkerspin12} and \cite{wkb}. For practical applications where an analytic solution of the KG equation is not available for $W_{\mathbf{p}} \equiv \int_{T}^{t} dt^{\prime} \ \omega_{\mathbf{p}}\left(t^{\prime}\right)$, a WKB solution such as the one described in \cite{wkb} is useful. One expands the solution as an asymptotic series in derivatives of the metric and truncates it at a given order. \\

The coefficients $a^{\mathbf{p}}_{\alpha s}$ and their adjoints are to be interpreted as annihilation and creation operators in the local inertial frame at $t = T$ after canonical quantisation. $u_{\alpha s}\left(\mathbf{p},t\right)$ carry spinor indices that have been suppressed for clarity and a label $s$ has been used in anticipation of the spin projection that will be defined in the next section. A general solution to~\eqref{eq:posfreq} can be written as
\begin{equation*}
\chi^{\alpha}_{b}\left(t\right) = \sum_{\alpha^{\prime}=-1}^{1}D^{\alpha}_{\alpha^{\prime}}\left(\mathbf{p}, t\right)A^{\mathbf{p}}_{\alpha^{\prime} s}\exp\left[-i\alpha W_{\mathbf{p}} \right]u_{\alpha s}
\end{equation*}
\\
$D_{\alpha\alpha^{\prime}}\left(\mathbf{p},t\right)$ are elements of the Bogoliubov matrix describing the mixing of positive and negative frequencies. The complete to~\eqref{eq:diracbianchi} can now be written as a superposition of plane waves
\small
\begin{equation}
\label{eq:planewavesoln}
\psi_{\alpha s} = \int \sqrt{-g} \frac{d^{3}\mathbf{p}}{\left(2\pi\right)^3} \sum_{\alpha^{\prime}}D_{\alpha\alpha^{\prime}}A_{\alpha^{\prime}s}^{\mathbf{p}}
u_{\alpha^{\prime}s}
\exp\left(-i\alpha^{\prime}W_{\mathbf{p}} + i\alpha\mathbf{p}\cdot\mathbf{x}\right)
\end{equation}\normalsize
where the full solution involves sums over $\alpha$ and $s$. The $A_{\alpha^{\prime}s}^{\mathbf{p}}$s and their adjoints are the coefficients written as a column vector
\begin{equation}
\label{eq:antcommutinit}
\vec{A}_{\alpha^{\prime}s^{\prime}}^{\mathbf{p}} = \begin{pmatrix}
a_{s^{\prime}}^{\mathbf{p}}\left(T\right) \\ b_{s^{\prime}}^{\dagger \mathbf{p}}\left(T\right)
\end{pmatrix} 
\end{equation}\\
To quantisatise the fermionic field, one imposes the anti-commutation conditions on them which results in the following relations between the coefficients
\begin{equation}\label{eq:anticomm2}
\left[a_{s}^{\mathbf{p}},a_{s^{\prime}}^{\mathbf{\dagger p^{\prime}}}\right]_{+} = \left[b_{s}^{\mathbf{p}},b_{ s^{\prime}}^{\dagger\mathbf{p^{\prime}}}\right]_{+}  = \delta_{ss^{\prime}}\delta_{pp^{\prime}}.
\end{equation} 
where the $T$ labels have been suppressed for clarity. The $a_{s^{\prime}}^{\mathbf{p}}\left(T\right)$ and $b_{s^{\prime}}^{\dagger \mathbf{p}}\left(T\right)$ can now be interpreted as annihilation and creation operators for a fermion and an anti-fermion of momentum $\mathbf{p}$ and spin $s^{\prime}$ respectively defined for the locally inertial frame at $t=T$. They define a $Q_{T}$ vacuum
\begin{equation}
\begin{split}
a_{s^{\prime}}^{\mathbf{p}}\left(T\right) \vert \ 0_{T} \ > \ & = \ 0 \\
b_{s^{\prime}}^{\mathbf{p}}\left(T\right) \vert \ 0_{T} \ > \ & = \ 0 \\
\end{split}
\end{equation}
which is empty with respect the local fermion and anti-fermion number operators \small $a_{s^{\prime}}^{\dagger\mathbf{p}}\left(T\right)a_{s^{\prime}}^{\mathbf{p}}\left(T\right)$ \normalsize and \small $b_{s^{\prime}}^{\dagger\mathbf{p}}\left(T\right)b_{s^{\prime}}^{\mathbf{p}}\left(T\right)$ \normalsize respectively. By applying the creation operators to the $Q_{T}$ vacuum, one generates a Fock space of locally defined particles and anti-particles as long as their wavelengths are not larger than the size of the local inertial frame in question. For longer wavelengths this approximate definition of particles and anti-particles is no longer valid. One therefore imposes a lower bound on the momentum, $\mathbf{p_{\mathbf{>}}}$, determined by requiring that the de Broglie wavelength of the particle, $\lambda_{B} = \frac{h}{\mathbf{p_{>}}}$, not be larger than the size of the region defined by setting equality in~\eqref{eq:killingcondbianchi1}. Writing factors of $\hbar$ and $c$ explicitly,
\begin{equation}\label{eq:localdefparticle}
\mathbf{p_{>}} = \frac{\hbar}{c\vert\mathcal{R}\left(T\right)\vert}.
\end{equation}

The creation and annihilation operators at an arbitrary time $t$ are
\begin{equation}\label{eq:timedeptopt}
\vec{a}_{s}^{\mathbf{p}}\left(t\right) = \mathbf{D}\left(\mathbf{p},t\right)\vec{A}_{s}^{\mathbf{p}}
\end{equation}
where $\mathbf{D}\left(\mathbf{p},t\right)$ is the Bogoliubov matrix. $\vec{a}_{s}^{\mathbf{p}}\left(t\right)$ must satisfy anti-commutation relations ~\eqref{eq:anticomm2} at all times, which is another way of saying that the matrix $\mathbf{D}\left(\mathbf{p},t\right)$ must describe canonical transformations. Direct substitution of~\eqref{eq:timedeptopt} into~\eqref{eq:anticomm2} shows that the following conditions must hold
\begin{align}
\vert D_{\alpha\alpha}\vert^{2} + \vert D_{\alpha-\alpha}\vert^{2} & = 1 \label{eq:canonopt1} \\
D_{\alpha\alpha}D^{\dagger}_{-\alpha\alpha} + D_{\alpha-\alpha}D_{-\alpha -\alpha}^{\dagger} & = 0 \label{eq:canonopt2}
\end{align}
Condition~\eqref{eq:canonopt1} can be implemented by normalising the Bogoliubov coefficients with the factor 
\begin{align*}
\mathcal{N} = \frac{1}{\sqrt{\vert D_{\alpha\alpha}\vert^{2} + \vert D_{\alpha-\alpha}\vert^{2}}}.
\end{align*}\\
Ultimately, one is interested in determining the number density of created (anti-)particles, \small $\vert D_{\alpha -\alpha} \vert^{2} $ \normalsize. To that end, a suitable form for $u_{\alpha^{\prime} s}\left(\mathbf{p},t\right)$ must be determined and then used them to find expressions for $D_{\alpha\alpha^{\prime}}\left(\mathbf{p},t\right)$. 
\subsubsection*{Pseudo-Helicity Operator}
A first step in attempting the two tasks mentioned above is to substitute the solution for a single wave of momentum $\mathbf{p}$ into~\eqref{eq:diracbianchi},
\small
\begin{equation}
\label{eq:bogdiffeq1}
\begin{split}
& \sum_{\alpha^{\prime}}\gamma_{0}\frac{dD_{\alpha\alpha^{\prime}}}{dt}u_{\alpha^{\prime}} = \sum_{\beta}D_{\alpha \beta}\exp\left(-i\left(\alpha^{\prime}-\beta\right)W_{\mathbf{p}}\right) \times \\ & 
\left[-\gamma_{0}\frac{d}{dt}u_{\beta} \ +  -i \left(\beta \omega_{\mathbf{p}}\gamma_{0} - \alpha\sum_{j=1}^{3} \frac{p_j}{r_{j}}\gamma_{j} - m\right)
u_{\beta}\right] 
\end{split}
\end{equation}\normalsize
where spin labels have been suppressed for clarity. \\
\\
Within any locally inertial frame, one should recover the Dirac equation in flat space-time. Inspection of~\eqref{eq:bogdiffeq1} shows that in the limit $D_{\alpha\beta}\left(\mathbf{p},t\right) = \delta_{\alpha\beta}$ and $u\left(\mathbf{p},t\right)$ independent of time - which are the appropriate limits for a locally inertial frame at an arbitrary time $t$ - the only surviving term is
\begin{equation}
\label{eq:bogdiffeq2}
\begin{split}
\left(\alpha  \omega_{\mathbf{p}}\gamma_{0} - \alpha\sum_{j=1}^{3} \frac{p_j}{r_{j}}\gamma_{j} - m\right)
u_{\alpha} = 0 
\end{split}
\end{equation}\\
This is the Dirac equation in flat space-time in terms of locally inertial four momentum - $u_{\alpha s}\left(\mathbf{p},t\right)$ must therefore be the Dirac spinor. However, a difficulty arises immediately. In flat space-time, the spinor is conveniently expressed in terms of two component helicity eigenstates. Helicity, which represents the component of the particle's spin along its direction of motion and is a conserved quantity in flat space-time, is defined by the operator  
\begin{equation}\label{eq:helicityoperator}
\Sigma_{\mathbf{p}}= \frac{1}{\vert \mathbf{p} \vert} \begin{pmatrix} \mathbf{\sigma}\cdot \mathbf{p} & 0 \\ 0 & \mathbf{\sigma}\cdot \mathbf{p} \end{pmatrix} 
\end{equation}\\
where the $\sigma_{j}$s are the Pauli matrices. In the most general curved space-time, neither the Hamiltonian nor the helicity operators represent conserved quantities. How then does one make sense of spin and helicity in curved space-time?   \\
\\
For the FLRW spacetime considered in \cite{parkerspin12}, the helicity operator commutes with the Hamiltonian as a consequence of the spatial isotropy of the metric. Inspection of~\eqref{eq:posfreq2} in the limit $r_{j} \rightarrow r \ \forall j$ shows that the commutation of the helicity operator with the Hamiltonian implies that the spinor remains an eigenvector of the helicity operator at all times. \\

For spacetime~\eqref{eq:bianchimetric}, the Hamiltonian operator is
\begin{equation}\label{eq:hamiltonian}
\mathcal{H} = \sum_{j=1}^{3} \alpha \frac{p_{j}}{r_{j}}\gamma_{0}\gamma_{j} - \frac{i}{2}V^{-1}V^{\prime} + m\gamma_{0}
\end{equation}
It is easily verified that the helicity operator~\eqref{eq:helicityoperator} does not commute with the Hamiltonian. However, the operator
\begin{equation}\label{eq:phelicityoperator}
\begin{split}
\Sigma^{P}_{\mathbf{p}} & = \frac{1}{\tilde{p}} \begin{pmatrix} \sum_{j=1}^{3}\sigma_{j}\frac{p_{j}}{r_{j}} & 0 \\ 0 & \sum_{j=1}^{3}\sigma_{j}\frac{p_{j}}{r_{j}} \end{pmatrix}
\end{split}
\end{equation}\\
does. Here $\tilde{p} \equiv \sqrt{\sum_{j=1}^{3} \left(\frac{p_{j}}{r_{j}}\right)^2}$. 

\subsubsection*{Dirac spinors in the pseudo-Helicity basis}
Eigenvectors of operator~\eqref{eq:phelicityoperator} are therefore particularly suitable for representing the Dirac spinor. I will call $\Sigma^{P}_{\mathbf{p}}$ the pseudo-Helicity operator as it is the flat space-time helicity operator in any locally inertial frame. In the FLRW limit, the pseudo-Helicity operator reduces to the helicity operator~\eqref{eq:helicityoperator} as the inverse scale factor \small $r^{-1}\left(t\right)$ \normalsize drops out of the matrix in~\eqref{eq:phelicityoperator} and is canceled by \small $\tilde{p}\left(t\right) = r\left(t\right)p$ \normalsize, with $p$ the magnitude of the (three) momentum vector. One can therefore have a meaningful notion of helicity over the entire manifold in FLRW space-times while in Bianchi Type I metrics of the form~\eqref{eq:bianchimetric}, helicity can only be well defined locally.  \\
\\
In choosing Dirac spinors for a general curved space-time, one therefore seeks solutions to~\eqref{eq:bogdiffeq2} with $u_{\alpha s}\left(\mathbf{p},t\right)$ eigenvectors of the pseudo-Helicty operator. These are 

\begin{equation}\label{eq:spinors}
\begin{split}
u_{\tiny{1}s} = N \begin{pmatrix}
\xi_{s} \\ \frac{s\tilde{p}}{\omega_{\mathbf{p}} + m}\xi_{s}
\end{pmatrix}, \ \ \ \ \ u_{\tiny{-1}s} = N \begin{pmatrix}
\frac{s\tilde{p}}{\omega_{\mathbf{p}} + m}\xi_{s} \\ \xi_{s}
\end{pmatrix}
\end{split}
\end{equation}  
where ant-fermion spinors are understood to have negative momentum, $N$ is a normalisation factor for the spinors, and 
\begin{equation}\label{eq:twocompspinor}
\begin{split}
\xi^{s} = \begin{pmatrix}
1 \\ f^{s}\left(p\right)
\end{pmatrix}, \ \ \ \ \ f^{s}\left(p\right) \equiv \frac{\frac{p_{1}}{r_{1}} + i\frac{p_{2}}{r_{2}}}{\frac{p_{3}}{r_{3}} + s\tilde{p}}
\end{split}
\end{equation}  
When the sign of the momentum is switched, $\xi_{s} \rightarrow \xi_{-s}$. Therefore, an anti-particle of spin $s$ should be understood as an anti-spinor with $\xi_{-s}\left(-p\right)$. As this does not affect calculations, it will not be made explicit to avoid cumbersome notation. It will be convenient to pick 
\begin{equation*}
\begin{split}
N = \sqrt{\frac{\omega_{\mathbf{p}} + m}{2m \ 2s \ \xi^{\dagger s}\xi^{s^{\prime}}}}, \ \ \ \ \
\xi^{\dagger s}\xi^{s^{\prime}} & = \frac{2s \tilde{p}}{\frac{p_{3}}{r_{3}} + s\tilde{p}} \ \delta_{ss^{\prime}}
\end{split}
\end{equation*} 
for later computations. With this normalisation, the spinors satisfy 
\begin{equation}\label{eq:spinornorm}
\bar{u}_{\alpha^{\prime} s^{\prime}}u_{\alpha s} = \delta_{\alpha^{\prime}\alpha}\delta_{ss^{\prime}}
\end{equation}
where $\bar{u}_{\alpha^{\prime} s} = u^{\dagger}_{\alpha^{\prime} s}\gamma_{0}$. 

\subsection{Bogoliubov Matrix Elements}
With this choice of spinors, equation~\eqref{eq:bogdiffeq1} for the Bogoliubov matrix elements now simplifies to 
\small
\begin{equation}
\label{eq:bogdiffeq3}
\begin{split}
& \sum_{\alpha^{\prime}}\gamma_{0}\frac{dD_{\alpha\alpha^{\prime}}}{dt}u_{\alpha^{\prime}} = \sum_{\beta}D_{\alpha \beta}\exp\left(-i\left(\alpha^{\prime}-\beta\right)W_{\mathbf{p}}\right)
\left(-\gamma_{0}\frac{d}{dt}u_{\beta} \right) 
\end{split}
\end{equation}
\normalsize
The orthogonality relation~\eqref{eq:spinornorm} can be used to isolate each component of the Bogoliubov matrix. Multiplying~\eqref{eq:bogdiffeq3} from the right by $u^{\dagger}_{\alpha^{\prime}s}$,
\begin{equation}
\label{eq:bogdiffeq4}
\begin{split}
\frac{dD_{\alpha\alpha^{\prime}}}{dt} = \sum_{\beta}D_{\alpha \beta}\theta_{\alpha \alpha^{\prime}\beta}\left(t\right)
\end{split}
\end{equation}
where
\begin{equation*}
\begin{split}
& \theta_{\alpha \alpha^{\prime}\beta}\left(t\right) \equiv -\alpha^{\prime}\exp\left(-i\left(\alpha^{\prime}-\beta\right) W_{\mathbf{p}} \right) \left(\bar{u}_{\left(\alpha^{\prime},s\right)}\frac{d}{dt}u_{\left(\beta,s\right)}\right) 
\end{split}
\end{equation*}
\normalsize
Integrating equation~\eqref{eq:bogdiffeq4}, one obtains a system of coupled linear Volterra integral equations of the second kind with kernel \small $\theta_{\alpha\alpha^{\prime}\beta}\left(t\right)$
\begin{equation}
\label{eq:intsol1}
\begin{split}
& D_{\alpha \alpha^{\prime}}\left(t\right) = D_{\alpha \alpha^{\prime}}\left(T\right) - \sum_{\beta} \int_{T}^{t} dt^{\prime} \
D_{\alpha \beta}\left(t^{\prime}\right)\theta_{\alpha\alpha^{\prime}\beta}\left(t^{\prime}\right) 
\end{split}
\end{equation}
\normalsize
In general, equation~\eqref{eq:intsol1} will not be solvable analytically. However, for times small compared to a characteristic value $\tau$ associated with the scale factors $r_{j}\left(t\right)$, a Neumann series solution can be written by using the inverse of $\tau$ as a small parameter. To see this, one considers a rescaled time $t \rightarrow \tau t$, which weighs each derivative by a factor of $\frac{1}{\tau} \equiv \lambda$. Since the kernel (see~\eqref{eq:kernelelements} below) contains only one time derivative, an iterative solution to~\eqref{eq:intsol1} can be written as a power series in $\lambda$:
\begin{equation}
\label{eq:intsol}
\begin{split}
D_{\alpha \alpha^{\prime}}\left(\eta\right) = D_{\alpha\alpha^{\prime}}^{0}\sum_{n=0}^{\infty}\lambda^{n}\left(
\int_{T}^{t} \ dt^{\prime} \ \theta_{\alpha \alpha^{\prime}}^{\alpha^{\prime}}\right)^{n} \\
+  D_{\alpha-\alpha^{\prime}}^{0}\sum_{n=1}^{\infty}\lambda^{n}\left(
\int_{T}^{t} \ dt^{\prime} \ \theta_{\alpha \alpha^{\prime}}^{-\alpha^{\prime}}\right)^{n}
\end{split}
\end{equation} 
\normalsize
where
\small
\begin{equation*}
\begin{split}
& \left(
\int_{T}^{t} \ dt^{\prime} \ \theta_{\alpha^{\prime}}^{\beta}\left(t^{\prime}\right)\right)^{0} = 1 \\
& \left(
\int_{T}^{t} \ dt^{\prime} \ \theta_{\alpha^{\prime}}^{\beta}\left(t^{\prime}\right)\right)^{n} = \int_{T}^{t_{n-1}} dt_{1}...dt_{n} \  \theta_{\alpha^{\prime}}^{\gamma_{1}}
\theta_{\gamma_{1}}^{\gamma_{2}}...\theta_{\gamma_{n-1}}^{\beta},
\end{split}
\end{equation*} \\
\normalsize and $\alpha$ labels have been suppressed for clarity. For the remainder of this work I will suppress powers of $\lambda$ to lighten the notation but it should be understood that the results are only valid for times equal to or smaller than a characteristic time associated with the scale factors $r_{j}\left(t\right)$. \\

$ D_{\alpha \alpha^{\prime}}^{0} \equiv D_{\alpha \alpha^{\prime}}\left(T\right)$ are the initial value of the Bogoliubov matrix elements. By construction, at the initial time $t = T$, the vacuum is empty and only pure states are present, implying 
\begin{equation}\label{eq:initconditions}
D_{\alpha \alpha^{\prime}}^{0} = \delta_{\alpha\alpha^{\prime}}
\end{equation}
Using the explicit form of the spinors given in~\eqref{eq:spinors}, the kernel elements $\theta_{\alpha \ \alpha^{\prime}\beta}\left(t\right)$ are
\begin{equation}\label{eq:kernelelements}
\begin{split}
\theta_{\alpha -\alpha \alpha}\left(t\right) & =  \frac{\alpha}{\omega_{\mathbf{p}}} \ \frac{1}{2\tilde{p}}\sum_{j=1}^{3} \left(\frac{p_{j}}{r_j}\right)^2 \frac{r_{j}^{\prime}}{r_{j}}\exp\left(i2\alpha W_{\mathbf{p}}\right)   \\
\\
\theta_{\alpha\alpha\alpha}\left(t\right) & = i \frac{\frac{p_{1}}{r_{1}}\frac{p_{2}}{r_{2}}}{\frac{p_{3}}{r_{3}} + s\tilde{p}} \frac{1}{2 \tilde{p}}\left( \frac{r_{1}^{\prime}}{r_{1}} - \frac{r_{2}^{\prime}}{r_{2}} \right)
\end{split}
\end{equation}
 
It can additionally be verified that 
\begin{equation}\label{eq:kernelsymmetry}
\begin{split}
\theta_{\alpha -\alpha \alpha}\left(t\right) & = - \ \theta_{\alpha \alpha -\alpha}\left(t\right) \\ 
\theta_{\alpha \alpha \alpha}\left(t\right)  & = - \ \theta_{\alpha -\alpha -\alpha}\left(t\right) 
\end{split}
\end{equation}
\\
Having obtained expressions for kernels of the Bogoliubov matrix elements, one naturally proceeds to find a closed form integral expression for them. It is found after some tedious algebra that the easiest way to make these Bogoliubov transformations canonical - having ~\eqref{eq:timedeptopt} obeying~\eqref{eq:anticomm2} - is to impose  $\theta_{\alpha\alpha\alpha}\left(t\right) = 0$, or $r_{1}\left(t\right) = r_{2}\left(t\right)$. There are less constraining ways of achieving canonical transformations, most notably by normalising the coefficients with the factor $\mathcal{N}$ mentioned above. However, as this diverts from the main subject treated in this work, I shall not consider them here and will henceforth set $r_{1}\left(t\right) = r_{2}\left(t\right)$, which result in the Bogoliubov matrix elements being simple trigonometric functions that automatically satisfy normalisation conditions ~\eqref{eq:canonopt1} and~\eqref{eq:canonopt2}. \\

Using~\eqref{eq:kernelsymmetry} and imposing time-ordering as is familiar from Quantum Field Theory, the Volterra integral solution for the Bogoliubov matrix elements are
\begin{equation}\label{eq:bogelementsfinal}
\begin{split}
D_{\alpha \alpha}\left(t\right) & = \cos\left(\Theta_{\alpha-\alpha\alpha}\left(t\right)\right) \\
D_{\alpha -\alpha}\left(t\right) & = \sin\left(\Theta_{\alpha-\alpha\alpha}\left(t\right)\right) 
\end{split}
\end{equation}
where $\Theta_{\alpha -\alpha\alpha}\left(t\right) = \int_{T}^{t} \theta_{\alpha -\alpha\alpha}\left(t\right)$. The trigonometric form of the solution is what one expects for fermionic Bogoliubov transformations. Two other properties are immediately verified
\begin{equation}\label{eq:bogelementsfinal2}
\begin{split}
D^{\dagger}_{\alpha\alpha^{\prime}}\left(t\right) & = D_{\alpha\alpha^{\prime}}\left(t\right)  \\
D_{\alpha\alpha^{\prime}}\left(t\right) & = \left(-1\right)^{\alpha^{\prime} - \alpha}D_{-\alpha-\alpha^{\prime}}\left(t\right) 
\end{split}
\end{equation}
and, from these, it can be seen that ~\eqref{eq:canonopt1} and~\eqref{eq:canonopt2} are satisfied, making the Bogoliubov transformations canonical. \\

Finally, the time-dependent number density - number of particles per unit volume at a time $t$ - of created fermions is 
\begin{equation}
\begin{split}
N\left(t\right) & = \sum_{s} \ < 0_T \left\vert a_{s^{\prime}}^{\dagger\mathbf{p}}\left(t\right)a_{s^{\prime}}^{\mathbf{p}}\left(t\right) \right\vert 0_{T} > \\
& =  2 \sin^{2}\left(\Theta_{\alpha-\alpha\alpha}\left(t\right)\right) 
\end{split}
\end{equation}
where the factor of 2 comes from the sum over spins. The number density of created anti-fermions is the same as the theory has no source of global $U\left(1\right)$ symmetry breaking. 

\subsection{Massless and Infinite Mass Limits}
I now come to the explicit proof that massless particles creation is possible in an anisotropic background space-time but vanishes when the isotropic limit is taken. As mentioned previously, several authors (see for example \cite{parkerspin0}, \cite{parkerspin12}, and \cite{tagirov}) have shown that no massless particles of non-zero spin - or spin zero particles but conformally coupled to the Ricci scalar - are created in an FLRW universe. Bronnikov and Tagirov, in particular, discussed the vacuum and the construction of Fock spaces for an FLRW space-time extensively in \cite{tagirov}. Their work considers only scalar fields conformally coupled to the Ricci scalar but the main argument applies to particles of non-zero spins as well: in a FLRW space-time and in the absence of a (conformal symmetry breaking) mass term a conformal time-like Killing vector and, by extension a $G$ vacuum, exists. To see this, it is best to first define two terms for a general metric. A conformal transformation is a local rescaling of the line element defined by 
\begin{equation}
g_{\mu\nu}\left(x\right)dx^{\mu}dx^{\nu} \rightarrow \exp\left[-2\Omega\left(x\right)\right]g_{\mu\nu}\left(x\right)dx^{\mu}dx^{\nu}
\end{equation} 
with $\Omega\left(x\right)$ a continuous, non-vanishing, finite, real function. A manifold $M$ with a metric $g_{\mu\nu}\left(x\right)$ is said to be conformally flat if there exists a space-time transformation which allows the line element to be recast in the form
\begin{equation}
g_{\mu\nu}\left(x\right)dx^{\mu}dx^{\nu} = H^{2}\left(x\right)\eta_{\mu\nu}dx^{\mu}dx^{\nu} 
\end{equation}
where $H\left(x\right)$ is an arbitrary function of space-time. Conformally flat metrics belong to the same equivalence class as the Minkowski metric and the former can be transformed into the latter through a conformal transformation. \\

From this, it is evident that the FLRW metric is conformally flat. The coordinate transformation $dt^{2} = r^{2}\left(t\right)d\eta^{2}$ that was used in Section~\eqref{ssec:b1} to show the presence of a conformal Killing vector casts the FLRW metric into a conformally flat form. This can now be better understood: the FLRW metric belonging to the same equivalence class as the Minkowski metric, the two are equivalent and have the same symmetries up to a conformal transformation. When the theory is conformally invariant, time translation is therefore a symmetry and a $G$ vacuum can be defined.\\

For the gravitational creation of massless fermions, one is concerned with how independent components of the anti-symmetric spin connection, $\omega_{\mu}^{\nu\rho}$, transform under a conformal transformation. For space-time~\eqref{eq:bianchimetric}, the spin connection is just the time derivative of the scale factor in a particular direction. Its non-zero independent components transform as 
\begin{equation}
\label{eq:spincontrans}
\omega_{j}^{j0} = r_{j}^{\prime}\left(t\right) \rightarrow  r_{j}^{\prime}\left(t\right) - \Omega^{\prime}\left(t\right)
\end{equation}
where $j = 1,2,3$ and $\Omega\left(x\right)$ is chosen to be a function of time only. Clearly, by picking 
\begin{equation*}
\Omega\left(t\right) = r_{j}\left(t\right)
\end{equation*} 
for a \textit{fixed} $j$ at least one component of the spin-connection for space-time~\eqref{eq:bianchimetric} can be eliminated. More than one component can be eliminated if the $r_j\left(t\right)$s are not all distinct. In particular, in the FLRW limit all three independent components are eliminated. \\

For the creation of massless particle creation in space-time~\eqref{eq:bianchimetric} assuming $r_{1} = r_{2}$, it is convenient to write the different scale factors as
\begin{equation}
\begin{split}
r_{1}\left(t\right) & = r_{2}\left(t\right) = r\left(t\right) \\
r_{3}\left(t\right) & = \sqrt{\delta^{2}\left(t\right) + r^{2}\left(t\right)}
\end{split}
\end{equation}
where $\delta\left(t\right)$ is a measure of the anisotropy, which need not be small.  Under a conformal transformation with $\Omega\left(t\right) = r\left(t\right)$, equation~\eqref{eq:kernelelements} for the non-zero kernel element is
\begin{equation}\label{eq:kernelelementmassless}
\begin{split}
\theta_{\alpha -\alpha \alpha}\left(t\right) & =  \frac{1}{2\tilde{p}^{2}} \frac{p_{3}^{2}}{1 + \left(\frac{\delta}{r}\right)^{2}} \frac{\delta^{\prime}\left(t\right)\delta\left(t\right)}{\sqrt{1+ \left(\frac{\delta}{r}\right)^{2}}}\exp\left(i2\alpha W_{\mathbf{p}}\right),   
\end{split}
\end{equation}
where it is assumed that $\tilde{p}$ and $W_{\mathbf{p}}$ are appropriately transformed. In the FLRW limit $\delta\left(t\right)=0$ and $\Theta_{\alpha -\alpha \alpha}\left(t\right) = 0$ - there is no massless fermion production, as anticipated. \\

Finally, it is easily shown, as a consistency check, that fermion creation stops in the infinite mass limit: $\theta_{\alpha -\alpha \alpha}\left(t\right)$ vanishes due to dependence on the inverse mass through the $\frac{1}{\omega_{p}}$ factor in~\eqref{eq:kernelelements}. This verifies the claim made previously that gravitational particle creation is more substantial for lighter particles - in the infinite mass limit, there is no particle creation, as one might expect intuitively.  

\section{Some Cosmological Implications}\label{sec:conc}
In this final section I mention some areas where massless gravitational particle creation might be consequential. Since the latter is not possible in isotropic universes, this is of particular interest for cosmological models with an early anisotropic phase. For a full computation one would need to evaluate the energy momentum tensor, most likely numerically using specific models of anisotropic gravitational fields such as the Krasner metric used in \cite{bianchi1spin0}. This formidable undertaking is left as a future work. Nonetheless, one can get a flavour of things to come by considering various approximations. \\

To estimate $\Theta_{\alpha-\alpha\alpha}\left(t\right)$, the exponential phase factor in~\eqref{eq:kernelelements} can be neglected since, from the Euler formula, this introduces an order one correction at most. In the massless limit and assuming $r_{1}\left(t\right) = r_{2}\left(t\right)$, 
\begin{equation}\label{eq:b1masslessub}
\Theta_{\alpha-\alpha\alpha}\left(t\right) \sim \frac{\alpha}{2}\log\left[\frac{\tilde{p}\left(t\right)}{\tilde{p}\left(T\right)}\right]
\end{equation}
where \small $\tilde{p}\left(t\right) = \sqrt{p_{1}^{2} + p_{2}^{2} + p_{3}^{2}\left(\frac{r_{3}\left(t\right)}{r\left(t\right)}\right)^{2}}$. \normalsize Here, conformal symmetry has been used to absorb $r_{1}\left(t\right) = r_{2}\left(t\right) = r\left(t\right)$ into a redefinition of the scale factors. In the FLRW limit, $r_{3}\left(t\right) = r\left(t\right)$, the right-hand side of~\eqref{eq:b1masslessub} is the logarithm of unity, which is zero - massless gravitational fermion creation is not possible. An estimate for energy density, $\epsilon$, of created fermions is
\begin{equation}
\epsilon \sim \int d^{3}p \ \tilde{p}\left(t\right) \ \sin^{2}\left[\frac{\alpha}{2}\log\left(\frac{\tilde{p}\left(t\right)}{\tilde{p}\left(T\right)}\right) \right]
\end{equation}
As before, I define $\delta^{2}\left(t\right) = r^{2}_{3}\left(t\right) - r^{2}\left(t\right)$ for convenience. Using the double angle formula and neglecting the resulting cosine term, which is at most an order one correction to the \small$\sim \Lambda^{4}$ \normalsize dependence of the energy density ($\Lambda$ is a momentum cutoff), 
\begin{equation*}
\epsilon \sim  \ 2\pi\frac{\Lambda^{4}}{\left(\hbar c\right)^3}\left[ \sqrt{\delta^{2}\left(t\right) +r^{2}\left(t\right)} + \frac{r\left(t\right)}{\delta\left(t\right)}\sinh^{-1}\left[\frac{\delta\left(t\right)}{r\left(t\right)}\right]\right]
\end{equation*}
where the $\hbar$ factor is made explicit. Clearly, for large anisotropies,
\begin{equation*}
\frac{\delta\left(t\right)}{r\left(t\right)} \gg 1,
\end{equation*}
the energy density of the gravitationally produced massless particles is substantial. This is of interest in situations where a large anisotropy results in a substantial backreaction on the metric, and for a quantitative estimate one must compare this energy density to that of the background gravitational field, $\epsilon_{G}$, obtained from the $00$ component of the Einstein tensor (without a cosmological constant term). Writing factors of $c$ and $G$ explicitly,
\begin{equation*}
\epsilon_{G} = \frac{c^4}{8\pi G}\left[\left(\frac{r^{\prime}\left(t\right)}{r\left(t\right)}\right)^{2} + 2 \ \frac{r^{\prime}\left(t\right)r_{3}^{\prime}\left(t\right)}{r\left(t\right)r_{3}\left(t\right)}\right]
\end{equation*}
For a measure of the back-reaction on the metric the ratio of the energy densities $\epsilon$ and $\epsilon_{G}$ must be found. To that end, one writes the momentum cutoff in terms of the dimensionless parameter $\gamma \geq 1$ defined previously as  $ T = \gamma t_{P}$ and obtains $\Lambda = \hbar T^{-1}$. The resulting ratio is
\begin{equation}\label{eq:edensitycomparison}
\frac{\epsilon}{\epsilon_{G}} \sim \ 8\pi^2 \times \left(\frac{t_{P}}{T}\right)^{2} \times \frac{\delta^{2}\left(t\right)r\left(t\right)}{r^{\prime}\left(t\right)\delta^{\prime}\left(t\right)}.
\end{equation}
At late initial times, $T \gg t_{P}$, the backreaction of the created fermions on the metric is small. However, as $T$ approaches $t_{P}$, the energy density is substantial and may cause backreaction that substantially alters the form of the metric. Even if one assumes $T \approx \left(10^{3}\right)t_{P}$, this suppression in~\eqref{eq:edensitycomparison} can be compensated with large anisotropies from the $\delta$ dependence of the same expression. Therefore, an early anisotropic phase and the resulting massless fermion creation may find applications to several problems of interest to Particle Physics and Cosmology, three of which I mention briefly here. \\

\textit{Dark Matter Production:} Gravitational particle creation as a mechanism for generating dark matter was discussed in \cite{superheavydm} and \cite{superheavydm2}, where the authors consider the production of very heavy particles in an FLRW universe. As shown in this work, this effect is less substantial for high mass particles, and massless particles cannot be created in FLRW universes. Hence, in FLRW universes one is left with the gravitational production of very massive particles, a process that is not efficient. \\

A better mechanism suggests itself if one considers that these masses are generated through a scalar acquiring a vacuum expectation value (VEV). The main result of this work is that in a  cosmological model with a very early anisotropic phase a substantial amount of massless particles will be produced. These can acquire a mass later as the scalar acquires a VEV. In addition to being more efficient, this is also less restrictive as there is no imposition on \textit{when} the VEV has to appear. In the FLRW case, since the particles have to be massive to be produced and production has to happen in the very early universe to be substantial, the phase transition for the appearance of the VEV is restricted to near Planck times. \\

\textit{Primordial Magnetic Fields:} Astrophysical observations indicate a lack of GeV range photon detection from astrophysical sources known to reach TeV level energies. The simplest explanation for this is the presence of magnetic fields which deflect the photons accelerated by inverse Compton scattering \cite{primbfield}. A compelling explanation that has received significant attention in the literature is that some physical process generates a \textit{seed} field which is then amplified by a dynamo mechanism - see \cite{primbfieldlarry} for a review. If one is focused on early universe solutions to the problem, the main difficulty to overcome is that photons, being massless, are not produced in an FLRW universe. The reason for this is exactly the same as for fermions discussed in this work: the presence of a time-like conformal Killing vector results in a $G$ vacuum and there is no gravitational photon creation. The approach most commonly taken in the literature to circumvent this problem has been to introduce conformal symmetry breaking interactions that enable the generation of magnetic fields. However, the conclusion of this section applies to photons as well: an early anisotropic phase would produce a potentially substantial amount of photons that may provide a seed magnetic field that is later amplified. The viability of an early anisotropic phase as a mechanism for \textit{primordial magnetogenesis} is left as a future work. \\

\textit{Anisotropic Cosmological Collapse}: Massless particle creation in Bianchi Type I space-times \textit{might} be of interest in the study of cosmological collapse. Classically, it is well established that if the Hawking-Penrose singularity theorems apply, any cosmological contracting phase ultimately results in the formation of a singularity. Zeldovich and Starobinsky discussed the subject briefly in \cite{bianchi1spin0}, arguing that the substantial scalar particle creation as the cosmological singularity is approach is unlikely to halt cosmological collapse or change the dynamics from a contracting phase to an expanding one. It is doubtful that this conclusion changes with fermions, even though the large energy density of created particles found in this work as the Planck time is approached will result in a similarly large degeneracy pressure. However, a quantitative study of anisotropic versus isotropic collapse in the presence of spin-$\frac{1}{2}$ fields is well motivated on account of the gravitational creation of massless fermions in the former case but not the latter.  \\

As a final note I would like to mention that even though this work has considered only the Bianchi Type I space-time, the main conclusion carries over to other cosmological scenarios such as large inhomogeneities in early times or FLRW space-times with inhomogeneous perturbations. As long as the cosmological model considered is not conformally flat massless particles will be created, with all the interesting consequences mentioned here.

\appendix

\section{Tetrad Formulation Of General Relativity}\label{sec:Tetradform}
The tetrad formulation of General Relativity has been used extensively in this work for the setting up of the quantisation method and subsequently in treating spin-$\frac{1}{2}$ fields in curved space-time. As this is not such a common topic, I will introduce it briefly in this Appendix. The interested reader is invited to see \cite{weinberg} for a thorough review. The following conventions are used: bars represent Lorentz indices, raised and lowered with the Minkowski metric $\eta_{\mu\nu}$, unbarred indices represent generally covariant indices, raised and lowered with $g_{\mu\nu}\left(x\right)$, and repeated indices of the same kind are summed over. \\

The prescription for incorporating gravity into a physical system is to replace all Lorentz tensors with objects that are covariant under general coordinate transformations. Mathematically, this means that in General Relativity the Poincar{\'e} symmetry group of Special Relativity is extended to the GL$\left(\mathbb{R},4\right)$ group. This approach works for Lorentz vectors and tensors but not for spinors as there are no representations of GL$\left(\mathbb{R},4\right)$ that transform like spinors under the Lorentz subgroup. To introduce spinors in curved space-time, one must use the tetrad formulation of General Relativity. \\ 

Under a coordinate transformation $x^{\mu}\rightarrow\xi^{\mu}\left(x\right)$, the line element transforms as 
\small
\begin{equation}
\begin{split}
g_{\mu\nu}\left(x\right)dx^{\mu}dx^{\nu} \rightarrow  g_{\mu\nu}\left(\xi\right)\frac{\partial\xi^{\rho}}{\partial x^{\mu}}\frac{\partial\xi^{\sigma}}{\partial x^{\nu}}dx^{\rho}dx^{\sigma} = \tilde{g}_{\mu\nu}\left(\xi\right)d\xi^{\mu}d\xi^{\nu}
\end{split}
\end{equation}
\normalsize
Invoking the equivalence principle, one can choose, at an arbitrary point $X$, a set of locally inertial coordinates $\xi^{\bar{\rho}}_{X}$ such that $\tilde{g}_{\mu\nu}\left(X\right) = \eta_{\mu\nu}$.  In other words,
\small
\begin{equation}
\begin{split}
g_{\mu\nu}\left(X\right)dx^{\mu}dx^{\nu} & =  \frac{\partial\xi^{\bar{\rho}}_{X}}{\partial x^{\mu}}\frac{\partial\xi^{\bar{\sigma}}_{X}}{\partial x^{\nu}}\vert_{x=X}\ \eta_{\bar{\rho}\bar{\sigma}}dx^{\bar{\rho}}dx^{\bar{\sigma}} \\
\frac{\partial\xi^{\bar{\sigma}}_{X}}{\partial x^{\nu}}\vert_{x=X} & \equiv e^{\bar{\rho}}_{\mu}\left(X\right)  
\end{split}
\end{equation}
\normalsize
The $e^{\bar{\rho}}_{\mu}\left(x\right)$ are called tetrads (or vierbiens) and can be raised and lowered as follows
\begin{equation}
e^{\bar{\rho}}_{\mu}\left(x\right) = \eta^{\bar{\rho}\bar{\sigma}}g_{\mu\nu}e^{\nu}_{\bar{\sigma}}\left(x\right)
\end{equation} 
Tetrads provide a local orthonormal basis for vectors. One can write any vector field $V^{\mu}\left(x\right)$ in terms of the coordinate system $\xi^{\bar{\rho}}_{X}\left(x\right)$ locally inertial at $x = X$,
\begin{equation}
\label{eq:vectetrad}
V_{X}^{\bar{\mu}}\left({x}\right) \equiv e^{\bar{\mu}}_{\nu}\left({x}\right)V^{\nu}\left({x}\right)
\end{equation}
where the subscript $X$ is used to indicate the fact that one has used a basis for the vectors that is locally orthonormal at the point $X$. 
\\
\\
The local inertial coordinates can be expanded in powers of $X$ \cite{blau}:
\small
\begin{equation}\label{eq:inertcoord}
\begin{split}
\xi^{\bar{\mu}}_{X}\left(x\right) & = \xi^{\bar{\mu}}_{X}\left(X\right) \ + \ \left(x^{\rho} - X^{\rho}\right) e^{\bar{\mu}}_{\rho}\left(X\right) \\ & + \left(x^{\rho} - X^{\rho}\right)\left(x^{\sigma} 
- X^{\sigma}\right)\Gamma_{\rho\sigma}^{\lambda}\left(X\right)e^{\bar{\mu}}_{\lambda}\left(X\right) + \mathcal{O}\left(X^{3}\right)
\end{split}
\end{equation}
\normalsize
The transformed metric, $\tilde{g}_{\mu\nu}\left(\xi\right)$, and Christoffel symbols, $\tilde{\Gamma}_{\mu\nu}^{\lambda}\left(\xi\right)$, in that local inertial frame are
\begin{equation}\label{eq:inertmetricchristof}
\begin{split}
\tilde{g}_{\mu\nu}\left(\xi\right) & = \eta_{\mu\nu} - \frac{1}{3}\tilde{R}_{\mu\rho\nu\sigma}\left(\xi_{0}\right) \left(\xi^{\rho} - \xi_{0}^{\rho}\right)\left(\xi^{\sigma} - \xi_{0}^{\sigma}\right) + \mathcal{O}\left(\xi^{3}\right) \\
\tilde{\Gamma}_{\mu\nu}^{\lambda}\left(\xi\right) & = -\frac{1}{3}\left(\xi^{\rho} - \xi_{0}^{\rho}\right)\left(\tilde{R}^{\lambda}_{\mu\nu\rho}\left(\xi_{0}\right) + \tilde{R}^{\lambda}_{\nu\mu\rho}\left(\xi_{0}\right) \right) + \mathcal{O}\left(\xi^{3}\right)
\end{split}
\end{equation}
where $x^{\mu}\left(\xi_{0}\right) = X^{\mu}$ and $\tilde{R}^{\lambda}_{\mu\nu\rho}\left(\xi_{0}\right)$ is the transformed Riemann tensor  
\begin{equation}\label{eq:riemanntsrlocal}
\tilde{R}_{\mu\rho\nu\sigma}\left(\xi_{0}\right) = \frac{\partial^2 g_{\mu\sigma}}{\partial x_{\rho}\partial x_{\nu}}\vert_{\xi = \xi_{0}} - \frac{\partial^2 g_{\mu\nu}}{\partial x_{\rho}\partial x_{\sigma}}\vert_{\xi = \xi_{0}} +\mathcal{O}\left(\xi^3\right)
\end{equation} 

\section{Spinors In Curved Space-time}\label{sec:DiracCrst}
Spin-$\frac{1}{2}$ particles in a background gravitational field are described by the covariant generalization of free Dirac equation, with the covariant derivative that includes a spin connection. A spinor $\psi\left(\mathbf{x},t\right)$ existing on a curved background space-time satisfies
\begin{equation}
\label{eq:diraccovariant}
i\left[ \ e^{\mu}_{\bar{\nu}}\gamma^{\bar{\nu}}D_{\mu} - m \ \right]\psi\left(\mathbf{x},t\right) = 0,
\end{equation}
with $D_{\mu}$ is the covariant derivative for spinor fields
\begin{equation}\label{eq:vierbiens}
\begin{split}
D_{\mu} & = \partial_{\mu} - \frac{i}{4}\omega^{\bar{\rho}\bar{\lambda}}_{\mu}\sigma_{\bar{\rho}\bar{\lambda}} \\
\sigma_{\bar{\rho}\bar{\lambda}} & = \frac{i}{2}\left[\gamma_{\bar{\rho}},\gamma_{\bar{\lambda}}\right]_{-} 
\end{split}
\end{equation}
The 4$\times$4 $\gamma$ matrices satisfy the Clifford Algebra
\begin{equation}
\left[\gamma^{\mu},\gamma^{\nu} \right]_{+} \equiv \gamma^{\mu}\gamma^{\nu} +\gamma^{\nu}\gamma^{\mu} = 2\eta^{\mu\nu} 
\end{equation}
Defining $\tilde{\gamma}^{\mu} \equiv e^{\mu}_{\bar{\nu}}\gamma_{\bar{\nu}}$,
\begin{equation}
\left[\tilde{\gamma^{\mu}},\tilde{\gamma^{\nu}} \right]_{+} = 2g^{\mu\nu}
\end{equation}
The $\sigma_{\bar{\rho}\bar{\lambda}}$ furnish a local spinor representation of the Lorentz group. The $\omega^{\bar{\rho}\bar{\lambda}}_{\mu}$ are the spin connection, anti-symmetric in $\bar{\rho}\bar{\lambda}$ exchange, defined in \cite{fujikawa} as:
\begin{equation}
\begin{split}
\omega^{\bar{\rho}\bar{\lambda}}_{\mu} & = \frac{1}{2}e^{\bar{\rho}\alpha}e^{\bar{\lambda}\beta}\left( C_{\alpha\beta\mu} - C_{\beta\alpha\mu} - C_{\mu\alpha\beta} \right) \\ 
C_{\alpha\beta\mu} & \equiv e^{\bar{\kappa}}_{\alpha}\left( \partial_{\beta}e_{\bar{\kappa}\mu} - \partial_{\mu}e_{\bar{\kappa}\beta} \right)
\end{split}
\end{equation}
 
\begin{acknowledgements}

I am especially grateful to Joseph Bramante for useful comments after proofreading this manuscript and for fruitful discussions about the applications of gravitational particle creation to dark matter freeze in and cosmology. I would also like to thank the two anonymous referees who provided thoughtful comments and useful references that have significantly improved this manuscript. This research was supported in part by Perimeter Institute for Theoretical Physics. Research at Perimeter Institute is supported by the Government of Canada through Industry Canada and by the Province of Ontario through the Ministry of Economic Development \& Innovation.

\end{acknowledgements}


\clearpage
\begin{thebibliography}{5}
\bibitem{parkerparticlecreation} L. Parker, Phys. Rev. Lett. \textbf{23}, 562 (1968).
\bibitem{parkerspin0} L. Parker, Phys. Rev. \textbf{183}, 1057 (1969)
\bibitem{parkerspin12} L. Parker, Phys. Rev. D\textbf{3}, 346 (1971)
\bibitem{dewitt} B. DeWitt, Center for Relativity, Department of Physics, University of Texas, Austin, Texas, USA, \textit{Quantum Field Theory in Curved Spacetime}, Phys. Rep. \textbf{19}, 6 (1975)
\bibitem{tagirov} E. A. Tagirov, N. A. Chernikov, Preprint JINR P2-3777, Dubna, 1968
\bibitem{bianchi1spin0} Y.B. Zeldovich, and A.A. Starobinsky, JETP \textbf{34}, 6, (1972)
\bibitem{weinberg}S. Weinberg, \textit{Gravitation and Cosmology: Principles and Applications of the General Theory of Relativity}, (John Wiley \& Sons, Inc., New York, 1972), pp. 365 - 373.
\bibitem{blau} M. Blau, \textit{Lecture Notes on General Relativity}, (Unpublished), pp. 91 - 99.
\bibitem{birrell} N. Birrell and P.Davies, \textit{Quantum Fields In Curved Space}, (Cambridge University Press, 1982)
\bibitem{parkerqftcurvedspacetime} L. Parker and D.Toms, \textit{Quantum Field Theory in Curved Spacetime: Quantized Fields and Gravity}, (Cambridge University Press, 2009).
\bibitem{fujikawa}K. Fujikawa and H. Suzuki, \textit{Path Integrals and Quantum Anomalies}, (Oxford University Press, 2013), pp. 253 - 258
\bibitem{superheavydm} D.J.H Chung, E.W. Kolb, and A. Riotto, Phys. Rev. D. \textbf{59}, 023501 (1998).
\bibitem{superheavydm2} V.A. Kuzmin, I.I Tkachev, JETP Lett. \textbf{68}, 4 (1998).
\bibitem{primbfield} S. Naoz and R. Narayan, Phys. Rev. Lett. \textbf{111}, 051303, (2013).
\bibitem{primbfieldlarry} L.M. Widrow, Rev. Mod. Phys. \textbf{74}, 775, (2002). 
\bibitem{woodard} R.P. Woodard,  Int. J. Mod. Phys. D \textbf{23}, 1430020, (2014).
\bibitem{blureview} B. L. Hu, arXiv: 1812.11851.
\bibitem{wkb} S. Habib, C. Molina-París, and E. Mottola, Phys. Rev. D \textbf{61}, 024010, (1999).
\end{thebibliography}
\end{document}